\documentclass[12pt]{article}
\def\slash#1{\setbox0=\hbox{$#1$}#1\hskip-\wd0\hbox to\wd0{\hss\sl/\/\hss}}
\usepackage{epsf, cite, amsmath, amssymb}
\usepackage{epsfig}
\usepackage[all]{xy}
\makeatletter
\renewcommand\section{\@startsection {section}{1}{\z@}%
                                   {-3.5ex \@plus -1ex \@minus -.2ex}%nn
                                    {2.3ex \@plus.2ex}% 
                                   {\normalfont\large\bfseries}}
\renewcommand\subsection{\@startsection{subsection}{2}{\z@}%
                                     {-3.25ex\@plus -1ex \@minus -.2ex}%
                                     {1.5ex \@plus .2ex}%
                                     {\normalfont\bfseries}}
\makeatother

\let\non\nonumber

\newcommand{\bea}{\begin{eqnarray}}
\newcommand{\eea}{\end{eqnarray}}
\newcommand{\be}{\begin{equation}}
\newcommand{\ee}{\end{equation}}

\newcommand{\p}{\partial}

\newcommand{\s}{\sigma}

\newcommand{\C}[1]{$(\ref{#1})$}

\begin{document}

\begin{titlepage}

\begin{center}

%\today
%\hfill                  

\hfill DAMTP-2006-19 

\vskip 2 cm
{\Large \bf Gravitino Propagator in anti de Sitter space}\\
\vskip 1.25 cm { Anirban Basu$^{a}$\footnote{email: abasu@ias.edu}
and Linda I. Uruchurtu $^{b}$\footnote{email: liu20@damtp.cam.ac.uk}
}\\
{\vskip 0.75cm
$^a$ Institute for Advanced Study, Princeton, NJ 08540, USA\\
\vskip 0.2cm
$^b$ Department of Applied Mathematics and Theoretical Physics,
Wilberforce Road, Cambridge,
CB3 OWA, UK\\
}

\end{center}

\vskip 2 cm

\begin{abstract}
\baselineskip=18pt

We construct the gauge invariant part of the propagator for the massless gravitino in $AdS_{d+1}$
by coupling it to a conserved current. We also derive the propagator for the massive 
gravitino. 

\end{abstract}

\end{titlepage}

\pagestyle{plain}
\baselineskip=18pt

\section{Introduction}

In order to formulate a theory of quantum gravity, it is necessary to understand the propagation
of particles in curved space--times. Calculation of correlation functions to measure scattering
amplitudes involves the propagators for various particles in this background geometry. So the 
calculation of propagators in curved space--times is a problem of physical interest. 
This also assumes significance following
the AdS/CFT correspondence~\cite{Maldacena:1997re,Gubser:1998bc,Witten:1998qj} which
proposes a duality between a theory of quantum gravity on $AdS_{d+1}$ and a 
$d-$dimensional conformal field theory living on its boundary. In fact, this correspondence
uses information about the theory of quantum gravity on $AdS_{d+1}$ to make predictions
about the boundary conformal field theory. In order to do so, one needs a prescription 
that relates the correlators in these two theories. The precise map between correlators 
in the two theories is obtained from the relation
schematically given by~\cite{Gubser:1998bc,Witten:1998qj}
\be \label{rulecal}
e^{-S [\phi_0]} = \langle e^{\int {\cal{O}} \phi_0} \rangle.\ee  
Here $S [\phi_0]$ stands for the string/M theory action evaluated on 
fields $\phi$ that take boundary values $\phi_0$, which act as sources for the operators
$\cal{O}$ in the boundary conformal field theory. This 
correspondence has been extensively studied
by considering supergravity backgrounds with choices of metric and 
fluxes that preserve the $SO(d,2)$ isometry of $AdS_{d+1}$ and analyzing the dual conformal 
field theory, and vice versa.       

So the calculation of bulk--to--bulk as well as bulk--to--boundary propagators for various
supergravity modes propagating in $AdS_{d+1}$ has been an active area of 
research (see ~\cite{Corley:1998qg,Volovich:1998tj,Matlock:1999fy}, and~\cite{D'Hoker:2002aw} for further
references.). Our aim is to 
calculate the bulk--to--bulk propagator for the massless as well as the massive 
gravitino. The kinetic term in the action for the gravitino in $AdS_{d+1}$
is given by
\be \label{eqn0}
{\cal{S}} = -\int d^{d+1} z {\sqrt{g}} 
{\bar\psi}_\mu \left( \Gamma^{\mu\nu\rho} D_\nu \psi_\rho 
+ m \Gamma^{\mu\nu} \psi_\nu\right), \ee
where
\be D_\mu \psi_\nu=\partial_\mu \psi_\nu 
+\frac{1}{8}\omega_{\mu}^{ab}[\gamma_a ,\gamma_b]
\psi_\nu -\Gamma_{\mu\nu}^{\rho}\psi_\rho.\ee

Note that the parameter $m$ is not the physical mass of the gravitino, but is related to it.
In fact the physically massless gravitino has non--zero $m$, as we shall discuss later. 
In this work, we are interested in constructing the propagator for the physically massless as well
as the physically massive gravitino. 

The massless gravitino in $AdS_{d+1}$ arises in theories that are obtained
from compactifications of supergravity that preserve at least some supersymmetry. 
So this gravitino 
lies in the massless BPS multiplet containing the graviton. In fact it is possible
to construct a theory of gravity possessing local supersymmetry only when the 
gravitino is massless.
This gravitino couples to the conserved supercurrent 
which follows by supersymmetry as the graviton couples to the conserved stress tensor. The
parameter $m$ is fixed which results in a gauge invariant action which we shall discuss
when we construct the massless gravitino propagator.

The massive gravitino arises in theories where
supersymmetry is broken, and the gravitino gets mass by the super Higgs mechanism. The
value of the parameter $m$ depends on the mechanism of supersymmetry breaking. Hence the
massive gravitino propagator is useful in computations in theories with broken supersymmetry
in $AdS_{d+1}$.   

On the other hand, the physical part of the massive Kaluza--Klein gravitino
propagator can be constructed 
along the lines of the massless gravitino propagator. This is because the massive 
KK gravitino in supersymmetric theories couples to a conserved 
current. Essentially it is due to the fact that the gravitino couples to a conserved
current in the parent theory, and though the choice of the vacuum 
configuration specified by the metric and fluxes spontaneously breaks the diffeomorphism invariance, 
the lower dimensional gravitino continues to couple to a current, which satisfies a generalized 
conservation equation. This conservation equation can be obtained directly from the conservation 
equation of the supercurrent in the parent theory, by analyzing the details of the KK compactification. 

We shall construct the generic massive gravitino propagator, as well as the massless gravitino 
propagator which couples to a conserved current. In order to calculate 
the propagator, we find it useful to proceed using 
the intrinsic geometric objects used in calculations in maximally symmetric spaces. 
The essential idea is to decompose the propagators given their tensor or spinor 
structures in a suitable basis defined by purely geometric objects with the 
coefficients given by functions of the geodesic distance between the two points, which
we review below.  
This approach was originally used to construct the bulk--to--bulk vector propagator 
in~\cite{Allen:1985wd} in maximally symmetric spaces by suitably decomposing 
bitensors in these spaces. For constructing the massless propagator, the number of physical 
structures gets reduced as we can 
neglect the pure gauge contributions. Using a different basis for the
decomposition, this has been used in~\cite{D'Hoker:1999jc} to deduce 
the gauge invariant part of the gauge boson and graviton propagators. The 
spinor propagator in four dimensions 
has been constructed in~\cite{Allen:1986qj} using geometric objects and has been shown 
to match with the result obtained previously~\cite{Burges:1985qq}.  
More recently, the calculation of 
the spinor propagator has been generalized to arbitrary dimensions 
in~\cite{Muck:1999mh}. We shall use this formalism to obtain the physical part of
the massless gravitino propagator in $AdS_{d+1}$ upto pure gauge 
contributions\footnote{See~\cite{Deser:2000de} for a discussion}.

We want to construct the gravitino propagator $\Theta_{\mu\nu'} (z,w)$ in 
$AdS_{d+1}$ defined by
\be 
\Theta_{\mu\nu'} (z,w) = \langle \psi_\mu (z) \bar\psi_{\nu'} (w)\rangle ,\ee
where $\mu$ and $\nu'$ refer to the points $z$ and $w$ respectively. 
Given the vector--spinor structure of the propagator, 
we need to define intrinsic objects in maximally symmetric 
spaces that carry vector and spinor indices. The vectorial objects are given 
by~\cite{Allen:1985wd} 
\be \label{vecder} n_\mu = D_\mu \mu (z,w), \quad n_{\nu'} = D_{\nu'} \mu (z,w),\ee 
where $\mu (z,w)$ is the geodesic distance between $z$ and $w$. One also
has the bitensor $g_{\mu\nu'} (z,w)$ under which vectors transform as
\be V_\mu (z) = g_\mu^{~\nu'} (z,w) V_{\nu'} (w).\ee
Also $n_\mu$ and $n_{\nu'}$ satisfy the relation  
\be n_\mu = -g_\mu^{~\nu'} n_{\nu'}. \ee
To account for the spinor indices, we have the bi--spinor parallel 
propagator~\cite{Allen:1986qj}
\be \label{spinpar}
\Lambda^\alpha_{~\beta'} (z,w),\footnote{As before, 
unprimed and primed indices refer to $z$
and $w$ respectively.}\ee      
under which spinors transform as 
\be \psi^\alpha (z) = \Lambda^\alpha_{~\beta'} (z,w) \psi^{\beta'} (w).\ee  
It is very convenient to decompose the gravitino propagator in terms of independent structures
constructed out of $n_\mu, n_{\nu'}, g_{\mu\nu'}$ and $\Lambda^\alpha_{~\beta'}$.
Thus the propagator can be written as
\bea \label{gravitinoprop}
\Theta_{\mu\nu'} (z,w) &=&  A_1 (\mu) g_{\mu\nu'} \Lambda + A_2 (\mu)
n_\mu n_{\nu'} \Lambda + A_3 (\mu) g_{\mu\nu'} n^\s \Gamma_\s \Lambda +
A_4 (\mu) n_\mu n_{\nu'} n^\s \Gamma_\s \Lambda \non \\ &&+ A_5 (\mu) 
n_\mu \Lambda \Gamma_{\nu'} + A_6 (\mu) \Gamma_\mu n_{\nu'}  
\Lambda + A_7 (\mu) n_\mu n^\s \Gamma_\s \Lambda \Gamma_{\nu'} 
\non \\ &&+ A_8 (\mu)
n_{\nu'} n^\s \Gamma_\s \Gamma_\mu \Lambda + A_9 (\mu) \Gamma_\mu \Lambda
\Gamma_{\nu'} + A_{10} (\mu) n^\s \Gamma_\s \Gamma_\mu \Lambda \Gamma_{\nu'}, \eea 
where the $A_i$'s are scalar functions of the geodesic distance $\mu$. 

Unlike the massive gravitino, for the massless gravitino, the structure 
in \C{gravitinoprop} can be simplified using
the gauge invariance of the theory, and one can write down the physical part
of the propagator using lesser number of independent structures, as we shall 
discuss later. The method of calculating the coefficients 
$A_i$ is the same as in~\cite{Anguelova:2003kf} apart from the overall
normalization, and so we shall be brief.  

In fact for the massless gravitino, in section 3 we shall show that these coefficients are given by
\bea  A_9 (y) = \lambda y^{-d/2} (y-1)^{-(d+1)/2}, \quad A_{10} = \sqrt{\frac{y-1}{y}} A_9, 
\qquad \qquad  \non
\\ A_6 = \sqrt{y(y-1)} \Big[ \frac{d A_9}{dy} -\frac{A_9}{2y} \Big] ,\quad
A_8 = \sqrt{y(y-1)} \Big[ \frac{d A_{10}}{dy} -\frac{A_{10}}{2(y -1)} \Big], \non \\
A_1 = -A_8 -(d+1) A_9, \quad A_2 = (d-1) A_8, \quad A_3 = -A_6 -(d+1) A_{10}, \non \\
A_4 = -2(d+1) A_{10} -(d+3) A_6, \quad A_5 = -A_6 -2 A_{10}, \quad 
A_7 = A_8, \eea 
where $2y= 1 + {\rm cosh} \mu$, and $\lambda$ is a constant which we fix. And finally
in section 4, we deduce the gauge invariant form of the gravitino propagator which is given 
by
\bea 
\Theta_{\mu\nu'} (z,w) &=&  A (u)
\p_\mu \p_{\nu'} u (\p u .\Gamma) \Lambda + B (u) \p_\mu u \p_{\nu'} u (\p u .\Gamma) \Lambda
\eea 
where 
\bea A (u) &=& -2^{d +3/2} \lambda (d+1) (u+1)  
u^{-(d+3)/2} (u+2)^{-d/2 -2}
, \non \\ 
B (u) &=& 2^{d + 3/2} \lambda (d+1) u^{-(d+5)/2} (u+2)^{-d/2 -3} \Big(
(d+2) (u+1)^2 +1 \Big) \eea

\section{The massive gravitino propagator}

We first consider the massive gravitino propagator for which the analysis is simpler than the
massless one. The analysis is similar to~\cite{Anguelova:2003kf}, to which we refer the reader
for various details. The main strategy is to substitute \C{gravitinoprop} into the equation 
of motion obtained from \C{eqn0} and solve for the coefficients. It is easier to solve
for the coefficients by simplifying the equation of motion along the lines 
of~\cite{Grassi:2000} which leads to the equivalent set of equations
\be D.\psi =0, \quad \Gamma . \psi =0, \quad (\slash{D} - m) \psi_\mu =0.\ee

Following the analysis of~\cite{Anguelova:2003kf}, we obtain a system of coupled differential 
equations involving only $A_9$ and $A_{10}$, which
leads to a differential equation for $A_9$. Changing variables to
\be 2y= 1 + {\rm cosh} \mu, \ee
and defining $A_9 = \sqrt{y} {\widetilde{A}}_9$, we have that 
\be \label{Heuneqn}
\Big[ y(y-1)(y-a) \frac{d^2}{dy^2} +\Big\{ 
(\alpha +\beta +1) y^2 
-\Big( \alpha +\beta +1 -\delta 
+a(\gamma +\delta) \Big)y +a \gamma \Big\} \frac{d}{dy}+(\alpha \beta y -q)\Big] 
{\widetilde{A}}_9 (y) =0,\ee
where
\bea \label{valpar}
&&a = \frac{(d+1)(d-1) -4m^2}{(d-1)^2 -4m^2} , \quad \alpha = \frac{d+1}{2} \pm m, 
\quad \beta = \frac{d+1}{2} \mp m, \quad \gamma = \delta = \frac{d+3}{2}, \non \\ 
&&q = \frac{(d-1)(d+1)(d^2 +2d +5) +8m^2 (2m^2 -d^2 - d -2)}{4[(d-1)^2 -4m^2]}.\eea

Now the differential operator in \C{Heuneqn} is the Heun operator, and so \C{Heuneqn} is 
solved by the Heun function (see~\cite{Heunbook} for various details). 
The choice of the specific Heun function to obtain the propagator in $AdS_{d+1}$ 
is as discussed in the
literature. The same issue arises, for example, in the choice of the hypergeometric function
in the expression for the scalar or the vector propagator. We want a solution in $AdS_{d+1}$
whose singularity
at $\mu =0$ has the same strength as in flat space, and which has the 
fastest fall--off at spatial
infinity corresponding to $\mu =\infty$~\cite{Allen:1985wd,Allen:1986qj}. While the 
former condition is natural as the two points approach each other, the later condition
is necessary in specifying boundary conditions and for the stability of $AdS$ against small 
fluctuations~\cite{Avis:1977yn,Breitenlohner:1982jf}.

The two solutions $H$ of Heun equation \C{Heuneqn} with the properties mentioned above
are given by 
\be \label{Heun1}
{\widetilde{A}}_9 (y) =
\lambda_\alpha y^{-\alpha} H \Big( \frac{1}{a}, 
\frac{q +\alpha[a(\alpha -\gamma -\delta +1) -\beta 
+\delta]}{a} ;\alpha, \alpha -\gamma +1, \alpha -\beta +1, \delta ;\frac{1}{y}
\Big),\ee
with characteristic exponent $\alpha$ and 
\be \label{Heun2}
{\widetilde{A}}_9 (y) =
\lambda_\beta y^{-\beta} H \Big(
\frac{1}{a}, \frac{q +\beta[a(\beta -\gamma -\delta +1) -\alpha 
+\delta]}{a} ; \beta -\gamma +1, \beta, \beta -\alpha +1, \delta ;\frac{1}{y}
\Big),\ee
with characteristic exponent $\beta$. Here $\lambda_\alpha$ and $\lambda_\beta$ are
arbitrary coefficients. Demanding the fastest fall-off at spatial infinity, considering 
$\alpha ,\beta$ with the upper signs from \C{valpar} we see that 
for $m >0$, \C{Heun1} is the required solution, while for $m <0$, \C{Heun2} is the required 
solution (with the situation reversed for $\alpha, \beta$ with the lower 
signs). These are the two 
local solutions of Heun equation at $y =\infty$ and have been obtained 
in~\cite{Maier} (see tables on pages 24 and 26).   

Now in order to fix the coefficient $\lambda$, one has to match the singularity 
as $\mu \rightarrow 
0$ (i.e., $y \rightarrow 1$) of the Heun function with the flat space answer. In order
to do this, one has to express the Heun function at $1/y$ in terms of 
Heun functions 
at $1-y$ and keep the leading term. However, unlike hypergeometric functions, 
for Heun functions such explicit expressions are not known in closed form in 
literature and so it is not easy to get an explicit expression for $\lambda$.
However, the coefficients relating two local solutions at two singular points have been 
implicity computed in~\cite{Heuncoeff}\footnote{We thank R. Maier for bringing this
reference to our notice.}. Using this, one can relate the local solution at $1/y$ to the ones
at $1-y$ and determine $\lambda$ implicitly.  

Though it is difficult to determine $\lambda$ in general, we can make a consistency check of 
\C{Heun1} and \C{Heun2}. Note that at $y=1$, the Heun 
function $H(a,q;\alpha,\beta,\gamma,\delta;y)$ has leading behaviour 
$(1-y)^{1-\delta}$~\cite{Heunbook} and so \C{Heun1} and \C{Heun2} both have leading
behaviour $\mu^{-(d+1)}$ as $\mu \rightarrow 0$ on using the expression for $\delta$ 
in \C{valpar}, thus leading to $A_9 \sim \mu^{-(d+1)}$ as $\mu \rightarrow 0$. 
Later on, we shall calculate the various coefficients for the flat space case
for the massless gravitino. Though the details of constructing the propagator are 
different for the massive and massless case, the scaling of the singularity is  
the same. In fact we shall see that in flat space $A_9 \sim \mu^{-(d+1)}$ as 
$\mu \rightarrow 0$ which agrees with the scaling deduced above.
It turns out that one can obtain the expressions for the remaining eight coefficients 
in \C{gravitinoprop} in terms of $A_9$ and $A_{10}$, where the equations involved
are purely algebraic~\cite{Anguelova:2003kf}. Thus this gives us the complete
expression for the massive gravitino propagator.

\section{The massless gravitino propagator}

In order to construct the massless gravitino propagator in $AdS_{d+1}$, we couple
the gravitino to a conserved current leading to the action   
\be \label{actgrav}
{\cal{S}} = -\int d^{d+1} z {\sqrt{g}} \left[ 
{\bar\psi}_\mu \left( \Gamma^{\mu\nu\rho} D_\nu \psi_\rho 
+ m \Gamma^{\mu\nu} \psi_\nu\right) + \left( {\bar\psi}_\mu {\cal{J}}^\mu + 
{\rm h.c.} \right) \right].\ee

The action for the massless gravitino has a gauge invariance because of which the gravitino
has the correct number of degrees of freedom~\cite{Deser:1977uq,Deser:1983mm}. In fact
the free part of the action \C{actgrav} has a gauge symmetry given 
by~\cite{Nolland:2000fx}\footnote{Local Weyl invariance has been used to identify a special
mass value for the scalar field in~\cite{Dorn:2003au}.}  
\be \label{gaugeinv}
\delta \psi_\mu \equiv {\mathcal{D}}_\mu \eta = D_\mu \eta -\frac{m}{d-1} 
\Gamma_\mu \eta ,\ee
for 
\be \label{valm}
m = \pm \frac{d-1}{2}.\ee
Thus the free action is gauge invariant, and hence the gravitino is massless,  
only for the specific values of $m$ given by \C{valm}. Thus demanding the interaction to be
gauge invariant leads to the covariant current conservation
\be \label{defconJ}
{\mathcal{D}}_\mu {\cal{J}}^\mu = \Big( D_\mu -\frac{m}{d-1} \Gamma_\mu
\Big) {\cal{J}}^\mu= 0.\footnote{Note that this is different from 
$D_\mu {\cal{J}}^\mu =0$ as naively expected.}\ee 

The general result for the massless gravitino \C{valm} 
is consistent with the results in literature, when 
one considers the KK spectrum of compactification of $d=11$ supergravity on $AdS_4 \times 
S^7$~\cite{Casher:1984ym}, $AdS_7 \times S^4$~\cite{Pilch:1984xy}; and type IIB 
supergravity on $AdS_5 \times S^5$~\cite{Kim:1985ez}.

We note that this feature of appearance of gauge invariance for massless particles is generic in
curved space--times. The analysis for the graviton has been performed 
in~\cite{Buchbinder:2000fy}, where only for
special values of the parameters in the action, the graviton is massless and the theory
has diffeomorphism invariance. 
It is also worth mentioning that in the discussion above as well as in the literature, a  
particle in curved space--time is defined to be massless if it has the same number of degrees of 
freedom as the massless particle in flat space--time. It does not necessarily mean that
the particle has null propagation, in fact, the propagation can have support inside the 
light--cone~\cite{Deser:1983mm}. However, it should be noted that it has been argued
that massless modes do have light--cone propagation in~\cite{Flato:1985ws}. 

We now need to solve for the coefficients in \C{gravitinoprop} for the massless 
gravitino propagator
\be \label{expprop}
\psi_\mu (z) = \int d^{d+1} w \sqrt{g} \Theta_{\mu\nu'} (z,w) {\cal{J}}^{\nu'} 
(w),\ee 
which we outline in some detail. 
We substitute \C{gravitinoprop} into the equation of motion 
obtained from \C{actgrav} given by
\be \label{eom}
\Gamma_\mu \slash{D} \Gamma .\psi - D_\mu \Gamma .\psi -\Gamma_\mu D.\psi 
+\slash{D} \psi_\mu + m\Gamma_{\mu\nu} \psi^{\nu} = -{\cal{J}}_\mu. \ee
As mentioned before, it is easier to obtain them after simplifying the equation of 
motion by obtaining expressions for the lower spin components in \C{eom} which are $D.\psi$ 
and $\Gamma.\psi$. In order to do this, 
we contract \C{eom} with $D^\mu$ and $\Gamma^\mu$ respectively, and use covariant
current conservation \C{defconJ} to obtain
\bea \slash{D} \slash{D} \Gamma .\psi -D^2 \Gamma .\psi -\slash{D} D.\psi 
+D^\mu \slash{D} \psi_\mu +m \Gamma^{\mu\nu} D_\mu \psi_\nu =-\frac{m}{d-1} 
\Gamma .{\cal{J}}, 
\non \\ (d-1) \slash{D} \Gamma .\psi -(d-1) D .\psi +dm \Gamma .\psi 
= -\Gamma .{\cal{J}} .\eea 
Then one can easily generalize the calculations of~\cite{Grassi:2000} to obtain the relations 
\be \label{rel1}
D .\psi = \frac{1}{d-1} \Gamma .{\cal{J}} , \quad \Gamma .\psi =0 , \ee
which when substituted in \C{eom} leads to the expression
\be \label{rel3}
(\slash{D} - m) \psi_\mu = -{\cal{J}}_\mu  
+ \frac{1}{d -1} \Gamma_\mu \Gamma .{\cal{J}}. \ee 

The system of equations \C{rel1} and \C{rel3} contain the same 
information as in
\C{eom}. So substituting \C{gravitinoprop} into the expressions 
\C{rel1} and \C{rel3} gives a coupled system of eighteen equations involving $A_1, \ldots, A_{10}$ and 
their first derivatives. These are given by equations (3.6)-(3.17) in~\cite{Anguelova:2003kf},
where $d=n-1$, and $A_1, \ldots, A_{10} = \alpha, \ldots, \kappa$, which can be seen by comparing \C{gravitinoprop}
and equation (3.5) in~\cite{Anguelova:2003kf}.
The homogeneous equations are obtained by setting the current to zero and have been solved in~\cite{Anguelova:2003kf}. 
We shall first find a solution to these equations in the massless limit which involves all the coefficients. 
Then we shall obtain the gauge invariant part of the propagator.

Setting the contact terms to zero and solving the system of equations gives us
the homogeneous solution which has an arbitrary coefficient which is independent of
$\mu$, and the role of the contact terms is to fix this coefficient. As discussed before, 
we need the solution in flat space including the overall normalization to fix the 
$AdS_{d+1}$ normalizatio. This is discussed in detail in Appendix A.

Now we proceed to obtain the coefficients in $AdS_{d+1}$. Clearly the homogeneous
system of equations is the same as in the massive case (the only difference is that $m$ is
fixed), and so we consider the solution for 
$A_9$ which is given by \C{Heuneqn}.
Now the Heun function has four singular points at $y=0,1,a$ and 
$\infty$~\cite{Heunbook}. From the value of $a$ in \C{valpar}, note that
for the class of solutions given by
\be \label{partcase}
d-1 = \pm 2m, \ee
the singular point $a$ coincides with the singular point at $\infty$ and so
the nature of the equation changes, as this is a degeneration limit. Now from \C{valm} we see
that \C{partcase} is precisely the case when the gravitino is massless.  

One should note that the expressions for the various other
coefficients in terms of $A_9$ are naively singular as they involve a factor 
of $(d-1)^2 -4m^2$ in the denominator, and the numerator also vanishes. So
one has to take care to obtain the coefficients, 
which are of course finite, in this degenerate limit. The procedure essentially
involves setting $d- 1 \pm 2m =\epsilon$ so that the singular points of the Heun
equation are at $0,1,O(1/\epsilon)$ and $\infty$ and then taking the limit $\epsilon 
\rightarrow 0$ smoothly at the end so that the singular point at $O(1/\epsilon)$
merges with the singular point at $\infty$. We will find the coefficients directly
by considering the equations they satisfy in the massless limit. We now present the
solutions for the coefficients below when $d-1 =2m$ is 
satisfied\footnote{The $d-1 =-2m$ case 
works analogously.}.      

One can directly deduce the equation satisfied by ${\widetilde{A}}_9$ in this limit
to get
\be \label{formA9}
\Big[ y(1-y) \frac{d^2}{dy^2} +\Big\{ \frac{d+3}{2} -(d+3) y   
\Big\} \frac{d}{dy} - (d+1) \Big] {\widetilde{A}}_9 (y) =0.\ee

Now the differential operator in \C{formA9} is the hypergeometric operator, and 
so the solution is a hypergeometric function, which has three singular points
at $y=0,1$ and $\infty$ as expected. The solution of \C{formA9} for
$AdS_{d+1}$ is as discussed before~\cite{Allen:1985wd,Allen:1986qj}, and so 
we need the local solution at $y =\infty$ with the fastest fall--off at spatial 
infinity. There are two such solutions to the hypergeometric equation
\be  y(1-y) \frac{d^2 u(y)}{dy^2} +[ \gamma -(\alpha +\beta +1) y ]
\frac{d u(y)}{dy} - \alpha\beta u(y) =0\ee      
given by~\cite{Abramowitz}
\be u(y) \sim y^{-\alpha} F \Big( \alpha, \alpha -\gamma +1, \alpha -\beta +1,
\frac{1}{y} \Big),
\quad u(y) \sim y^{-\beta} F \Big( \beta, \beta -\gamma +1, \beta -\alpha +1,
\frac{1}{y} \Big).\ee

From \C{formA9} we see that 
\be \gamma =\frac{d+3}{2}, \quad (\alpha, \beta) = (d+1,1) \quad{\rm or} 
\quad (\alpha, \beta) = (1,d+1).\ee

Considering the solution with the fastest fall--off at spatial infinity we get that
\bea \label{expA91}
A_9 (y) &=& \lambda y^{-d-1/2} F \Big( d+1,\frac{d +1}{2};d+1;
\frac{1}{y} \Big) \non \\
&=& \lambda y^{-d/2} (y-1)^{-(d+1)/2}.\eea

To obtain the expression for $\lambda$, we compare it to the flat space 
expression given by \C{expA92}.
Equating the leading singularity in \C{expA91} and \C{expA92} as $\mu 
\rightarrow 0$, we see 
that 
\be \label{vallambda}
\lambda = -\frac{2^{(d+1)/2} \Gamma (\frac{d+1}{2}) C}{(d-1)^{d+2}} 
= -\frac{\Gamma(\frac{d-1}{2})}{(4\pi)^{(d+1)/2} (d+1)}, \ee
on using \C{valC}.

Having obtained $A_9$, we note that $(A_9, A_{10})$ form a coupled system of 
equations~\cite{Anguelova:2003kf}. In fact knowing $A_9$, we see that $A_{10}$ 
can be solved
algebraically using the equation
\bea \label{exprA10}
&&m \Big[ 4m^2 -(d-1)^2 +\frac{2(d-1)}{y-1} \Big] A_{10} = [4m^2 -(d-1)^2] 
\sqrt{y(y-1)}  \frac{d A_9}{dy} \non \\ &&+ \Big[ \sqrt{\frac{y-1}{y}} \Big( 
\frac{(d-2)(4m^2 -(d-1)^2)}{2} -(d-1)^2 \Big) + 4m^2 \sqrt{\frac{y}{y-1}}\Big] A_9.    
\eea
For the case $2m = d-1$, \C{exprA10} simplifies and gives
\be \label{valA10}
A_{10} = \sqrt{\frac{y-1}{y}} A_9 = \lambda y^{-(d+1)/2} (y-1)^{-d/2}.\ee

We can now
write down the expressions for $A_6$ and $A_8$ in terms of $A_9$ and $A_{10}$. One
can show that $A_6$ satisfies the equation~\cite{Anguelova:2003kf} 
\be \frac{A_6}{y} = \frac{d A_{10}}{dy} +\frac{(d-2) A_{10}}{2(y-1)} 
-\frac{m A_9}{\sqrt{y(y-1)}}. \ee
Expressing $A_{10}$ in terms of $A_9$ and working in this limit gives
\be A_6 = \sqrt{y(y-1)} \Big[ \frac{d A_9}{dy} -\frac{A_9}{2y} \Big]. \ee

Similarly using the equation for $A_8$ given by
\be \frac{A_8}{y-1} = \frac{d A_9}{dy} +\frac{(d-2) A_9}{2y} 
-\frac{m A_{10}}{\sqrt{y(y-1)}}, \ee
gives
\be A_8 = \sqrt{y(y-1)} \Big[ \frac{d A_{10}}{dy} -\frac{A_{10}}{2(y -1)} \Big]. \ee

Having obtained $(A_6,A_8,A_9,A_{10})$, the other coefficients are given by
\bea  \label{relcoeff}
A_1 = -A_8 -(d+1) A_9, \quad A_2 = (d-1) A_8, \quad A_3 = -A_6 -(d+1) A_{10}, \non \\
A_4 = -2(d+1) A_{10} -(d+3) A_6, \quad A_5 = -A_6 -2 A_{10}, \quad 
A_7 = A_8, \eea
thus completing the construction of the propagator.

As mentioned before, the propagator constructed above can be rewritten using the gauge 
invariance in the theory. This simplifies the calculation of 
Witten diagrams involving the massless
gravitino exchange in $AdS_{d+1}$. We should mention that in calculating the gravitino propagator 
we have worked on the subspace of conserved currents and did not use any gauge--fixing to obtain the 
propagator. We now proceed to obtain the gauge invariant part of the propagator.

\section{The gauge invariant massless gravitino propagator}
 
The propagator in any theory which has a gauge invariance is ambiguous. One always has the freedom
to add pure gauge terms to the propagator, which does not affect physical gauge--invariant 
amplitudes. So using this freedom, one can write the propagator in a convenient form.   
The propagator \C{gravitinoprop} has ten independent structures which follows from its 
vector--spinor structure, while the coefficients are determined by the equation of motion.
Now one can use the gauge invariance of the theory to add appropriately chosen 
pure gauge terms so that one has less than the ten independent structures in the expression for 
the propagator. We shall construct the propagator which has the least number of the independent 
structures in \C{gravitinoprop}, which we call the gauge invariant part of the propagator. 
Any other 
expression for the propagator will differ from the one we construct by pure gauge terms.  
The pure gauge terms we shall add to \C{gravitinoprop}  
are of the form ${\cal{D}}_\mu (z) P_{\nu'} (z,w)$ or ${\cal{D}}_{\nu'} (w) 
Q_\mu (z,w)$\footnote{We consider the case $2m = d-1.$}
with suitable choices for $P_{\nu'}$ and $Q_\mu$.  
This is because ${\cal{D}}_\mu (z) P_{\nu'} (z,w) $ yields a pure gauge term of the form
\be \psi_{\mu}^{\rm{pure~gauge}} (z) =  {\cal{D}}_\mu (z) \int d^{d+1} w \sqrt{g} 
P_{\nu'} (z,w) {\cal{J}}^{\nu'} (w) ,\ee
which can be removed by a gauge transformation using \C{gaugeinv}, and 
${\cal{D}}_{\nu'} (w) Q_\mu (z,w)$ yields
\be \psi_{\mu}^{\rm{pure~gauge}} (z) = -\int d^{d+1} w \sqrt{g} Q_\mu (z,w) 
{\cal{D}}_{\nu'} (w) {\cal{J}}^{\nu'} (w),\ee 
which vanishes due to current conservation using \C{defconJ}. We will consider only
those $P_{\nu'}$ and $Q_\mu$ which vanish as $u \rightarrow \infty$, so that the boundary 
contributions can be neglected. 

Given the index structure of the propagator, the 
only non--trivial possibilities that are pure gauge contributions are
\bea \label{setzero}
&&{\cal{D}}_\mu \Big( F_1 (u) n_{\nu'} \Lambda \Big), \quad  {\cal{D}}_\mu \Big( F_2 (u) \Lambda 
\Gamma_{\nu'} \Big) ,
\quad {\cal{D}}_\mu \Big( F_3 (u) n_{\nu'} n^\s \Gamma_\s \Lambda \Big) , 
\non \\&& {\cal{D}}_\mu \Big( F_4 (u) n^\s
\Gamma_\s \Lambda \Gamma_{\nu'} \Big) , \quad
{\cal{D}}_{\nu'} \Big( F_5 (u) n_\mu \Lambda \Big), \quad  {\cal{D}}_{\nu'} \Big( F_6 (u) 
\Gamma_\mu \Lambda \Big) ,
\non \\
&& {\cal{D}}_{\nu'} \Big( F_7 (u) n_\mu  \Lambda n^{\s'} \Gamma_{\s'} \Big) , \quad  
{\cal{D}}_{\nu'} \Big( F_8 (u)
\Gamma_\mu  \Lambda n^{\s'} \Gamma_{\s'} \Big), 
\eea 
where the $F_i$'s are arbitrary functions of $u$. Note that naively one might 
also want to add four terms  
${\cal{D}}_\mu \Big( F_9 (u) g_{\nu' \s}  \Gamma^\s \Lambda \Big)$, ${\cal{D}}_\mu 
\Big( F_{10} (u) g_{\nu' \s} n^\lambda
\Gamma_\lambda \Gamma^\s \Lambda \Big)$, ${\cal{D}}_{\nu'} \Big( F_{11} (u) g_{\mu \s'}  
\Lambda \Gamma^{\s'} 
\Big)$ and ${\cal{D}}_{\nu'} \Big( F_{12} (u) g_{\mu \s'} n^\lambda
\Gamma_\lambda \Lambda \Gamma^{\s'}\Big)$ to \C{setzero} . However using the relations  
\be n_\mu = -g_\mu^{~\nu'} n_{\nu'} , \qquad \Gamma_\mu \Lambda 
= g_\mu^{~\nu'} \Lambda \Gamma_{\nu'},\ee 
we see that they give back terms in \C{setzero} and thus are not linearly independent. 
We find it convenient to express the scalar functions in terms of the chordal distance $u$ 
defined by $u = {\rm cosh} \mu -1$. In fact, the basic vectorial and bitensor 
structures are given 
in terms of the chordal distance by
\be \label{changebas2} n_\mu =\frac{\p_\mu u}{\sqrt{u(u+2)}}, \quad
\quad n_{\nu'} =\frac{\p_{\nu'} u}{\sqrt{u(u+2)}}, \quad
g_{\mu\nu'} = -\p_\mu \p_{\nu'} u +\frac{\p_\mu u \p_{\nu'} u}{u+2}.\ee 

We now obtain expressions for the pure gauge terms given by \C{setzero} in terms of 
the various structures in \C{gravitinoprop}. To do so, we use the relations
\bea 
D_\mu n_{\nu'} = -\frac{1}{\sqrt{u(u+2)}} (g_{\mu\nu'} + n_\mu n_{\nu'}) ,\non \\
D_{\mu'} n_\nu = -\frac{1}{\sqrt{u(u+2)}} (g_{\mu'\nu} + n_{\mu'} n_\nu) ,\non \\
D_\mu n_\nu = \frac{u+1}{\sqrt{u(u+2)}} (g_{\mu\nu} - n_\mu n_{\nu}), \non \\
D_{\mu'} n_{\nu'} = \frac{u+1}{\sqrt{u(u+2)}} (g_{\mu'\nu'} - n_{\mu'} n_{\nu'}), 
\non \\ D_\mu \Lambda = \frac{1}{2} \sqrt{\frac{u}{u+2}} (\Gamma_\mu
\Gamma^\nu n_\nu -n_\mu) \Lambda, \non \\D_{\mu'} \Lambda = -\frac{1}{2} \sqrt{\frac{u}{u+2}}
\Lambda (\Gamma_{\mu'} \Gamma^{\nu'} n_{\nu'} - n_{\mu'} ).
\eea

Straightforward calculation gives 
\bea \label{group1}
{\cal{D}}_\mu \Big( F_1 (u) n_{\nu'} \Lambda \Big) &=&\frac{F_1}{\sqrt{u(u+2)}} \Big[
\Big( u(u+2) \frac{F_1'}{F_1} -1 -\frac{u}{2} \Big) n_\mu n_{\nu'} \Lambda - g_{\mu\nu'} 
\Lambda +\frac{u}{2}
n_{\nu'} \Gamma_\mu n . \Gamma \Lambda \Big]  \non \\
%&&=\frac{F_1}{\sqrt{u(u+2)}} \Big[
%\Big( u(u+2) \frac{F_1'}{F_1} -1 +\frac{u}{2} \Big) n_\mu n_{\nu'} \Lambda - g_{\mu\nu'} 
%\Lambda -\frac{u}{2}
%n_{\nu'} n^\s \Gamma_\s \Gamma_\mu \Lambda \Big] 
&&-\frac{F_1}{2} n_{\nu'} \Gamma_\mu \Lambda , \non \\
{\cal{D}}_{\nu'} \Big( F_5 (u) n_\mu \Lambda \Big) &=& \frac{F_5}{\sqrt{u(u+2)}} \Big[
\Big( u(u+2) \frac{F_5'}{F_5} -1 -\frac{u}{2} \Big) n_\mu n_{\nu'} 
\Lambda - g_{\mu\nu'} \Lambda -\frac{u}{2}
n_{\mu} n .\Gamma \Lambda \Gamma_{\nu'} \Big] \non \\
&&-\frac{F_5}{2} n_\mu \Lambda \Gamma_{\nu'}, \non \\
{\cal{D}}_\mu \Big( F_4 (u) n .
\Gamma \Lambda \Gamma_{\nu'} \Big)  &=& F_4 {\sqrt{\frac{u+2}{u}}} \Big[
\Big( \frac{u F_4'}{F_4} -\frac{1}{2} ) n_\mu n . \Gamma \Lambda \Gamma_{\nu'} 
+\frac{1}{2} \Gamma_\mu \Lambda \Gamma_{\nu'} \Big] 
\non  \\ &&-\frac{F_4}{2} \Gamma_\mu n.\Gamma \Lambda 
\Gamma_{\nu'}, \non \\
{\cal{D}}_{\nu'} \Big( F_8 (u)
\Gamma_\mu  \Lambda n' . \Gamma' \Big) &=& - F_8 {\sqrt{\frac{u+2}{u}}} 
\Big[ \Big( \frac{u F_8'}{F_8} -\frac{1}{2} \Big)  
n_{\nu'} \Gamma_\mu n . \Gamma \Lambda  -\frac{1}{2} \Gamma_\mu \Lambda
\Gamma_{\nu'} \Big]  \non \\ &&+\frac{F_8}{2} \Gamma_\mu n. \Gamma \Lambda \Gamma_{\nu'}, 
\non \\
%&& =- F_8 {\sqrt{\frac{u+2}{u}}} 
%\Big[ \Big( \frac{2 u F_8'}{F_8} -1 \Big) n_\mu n_{\nu'} \Lambda
%-\Big( \frac{u F_8'}{F_8} -\frac{1}{2} \Big) n_{\nu'} n^\s \Gamma_\s 
%\Gamma_\mu\Lambda  -\frac{1}{2} \Gamma_\mu \Lambda
%\Gamma_{\nu'} \Big], \non \\ 
{\cal{D}}_\mu \Big( F_2 (u) \Lambda 
\Gamma_{\nu'} \Big)  &=& \frac{F_2}{\sqrt{u(u+2)}}
\Big[ \Big( u(u+2) \frac{F_2'}{F_2} -\frac{u}{2} \Big) 
n_\mu \Lambda \Gamma_{\nu'} +\frac{u}{2} 
\Gamma_\mu
n . \Gamma \Lambda \Gamma_{\nu'}\Big] \non \\&&
-\frac{F_2}{2} \Gamma_\mu \Lambda \Gamma_{\nu'}, \non \eea
%&&= \frac{F_2}{\sqrt{u(u+2)}}
%\Big[ \Big( u(u+2) \frac{F_2'}{F_2} +\frac{u}{2} \Big) 
%n_\mu \Lambda \Gamma_{\nu'} -\frac{u}{2} 
%n^\s \Gamma_\s \Gamma_\mu \Lambda \Gamma_{\nu'}\Big], \non \\
\bea
{\cal{D}}_{\nu'} \Big( F_6 (u) 
\Gamma_\mu \Lambda \Big) &=& \frac{F_6}{\sqrt{u(u+2)}}
\Big[ \Big( u(u+2) \frac{F_6'}{F_6} -\frac{u}{2} \Big) 
n_{\nu'} \Gamma_\mu 
\Lambda  -\frac{u}{2} \Gamma_\mu  n . \Gamma
\Lambda \Gamma_{\nu'} \Big] %&&= \frac{F_6}{\sqrt{u(u+2)}}
%\Big[ \Big( u(u+2) \frac{F_6'}{F_6} -\frac{u}{2} \Big) 
%n_{\nu'} \Gamma_\mu \Lambda -u n_\mu \Lambda \Gamma_{\nu'}
%+\frac{u}{2} n^{\s} \Gamma_{\s} \Gamma_\mu
%\Lambda \Gamma_{\nu'} \Big] , 
\non \\
&& -\frac{F_6}{2} \Gamma_\mu \Lambda \Gamma_{\nu'} ,\non \\
{\cal{D}}_\mu \Big( F_3 (u) n_{\nu'} n . \Gamma \Lambda \Big) &=&
\frac{F_3}{\sqrt{u(u+2)}} \Big[ 
\Big( u(u+2) \frac{F_3'}{F_3}  -\frac{u+4}{2} \Big) n_\mu n_{\nu'} n .\Gamma 
\Lambda \non \\ &&- g_{\mu\nu'}
n . \Gamma \Lambda 
+\frac{u+2}{2} n_{\nu'} \Gamma_\mu \Lambda \Big] -\frac{F_3}{2} \Gamma_\mu n_{\nu'}
n .\Gamma \Lambda , \non \\
{\cal{D}}_{\nu'} \Big( F_7 (u) n_\mu  \Lambda n' . \Gamma' \Big) &=&
-\frac{F_7}{\sqrt{u(u+2)}} \Big[ 
\Big( u(u+2) \frac{F_7'}{F_7}  -\frac{u+4}{2} \Big) n_\mu n_{\nu'} n . \Gamma 
\Lambda \non \\ &&- g_{\mu\nu'}
n . \Gamma \Lambda 
-\frac{u+2}{2} n_\mu \Lambda \Gamma_{\nu'} \Big] +\frac{F_7}{2} n_\mu n .\Gamma 
\Lambda \Gamma_{\nu'} , \eea
where $F_i'$ denotes derivative with respect to $u$. 

We now show using \C{group1} that it is possible to express the propagator \C{gravitinoprop}
in terms of only two of the ten independent structures. We begin by rewriting
\C{gravitinoprop} as
\bea \label{gravitinoprop2}
\Theta_{\mu\nu'} (z,w) &=&  A_1  g_{\mu\nu'} \Lambda + (A_2 + 2 A_8 )
n_\mu n_{\nu'} \Lambda + A_3  g_{\mu\nu'} n . \Gamma \Lambda +
A_4 n_\mu n_{\nu'} n . \Gamma \Lambda \non \\ &&+ A_6 ( \Gamma_\mu n_{\nu'}  
\Lambda - n_\mu \Lambda \Gamma_{\nu'})+ A_8 ( n_\mu n . \Gamma  \Lambda \Gamma_{\nu'} 
- n_{\nu'} \Gamma_\mu n . \Gamma \Lambda) \non \\ &&+ A_9 \Gamma_\mu \Lambda
\Gamma_{\nu'} - A_{10} \Gamma_\mu n .  \Gamma \Lambda \Gamma_{\nu'}, \eea
where we have used the relations among the $A_i$'s. Now using the relations involving 
$F_2$ and $F_6$ from \C{group1}, we see that 
\bea &&A_6 \Big( \Gamma_\mu n_{\nu'}  
\Lambda - n_\mu \Lambda \Gamma_{\nu'} \Big) = 
{\cal{D}}_{\nu'} \Big( \sqrt{u+2} F \Gamma_\mu \Lambda \Big) \non \\&&
- {\cal{D}}_\mu \Big( \sqrt{u+2} F \Lambda \Gamma_{\nu'} \Big) +\sqrt{u} F \Gamma_\mu n .\Gamma
\Lambda \Gamma_{\nu'}
\eea
where 
\be F' = \frac{A_6}{\sqrt{u} (u+2)}.\ee
Similarly using the relations involving 
$F_4$ and $F_8$ from \C{group1}, we see that 
\bea &&A_8 \Big( n_\mu n. \Gamma \Lambda \Gamma_{\nu'}  - n_{\nu'} \Gamma_\mu n .\Gamma
\Lambda \Big) = 
{\cal{D}}_{\mu} \Big( \sqrt{u} G n. \Gamma \Lambda \Gamma_{\nu'} \Big) 
\non \\ &&+ {\cal{D}}_{\nu'} \Big( \sqrt{u} G  \Gamma_\mu \Lambda n' . \Gamma' \Big) 
-\sqrt{u+2} G \Gamma_\mu 
\Lambda \Gamma_{\nu'},
\eea
where 
\be G' = \frac{A_8}{u \sqrt{u+2}}.\ee

Thus neglecting the ${\cal{D}}_\mu (\ldots)$ and ${\cal{D}}_{\nu'} (\ldots)$ terms, 
we get that
\bea \label{gravitinoprop3}
\Theta_{\mu\nu'} (z,w) &=&  A_1  g_{\mu\nu'} \Lambda + (A_2 + 2 A_8 )
n_\mu n_{\nu'} \Lambda + A_3  g_{\mu\nu'} n . \Gamma \Lambda +
A_4 n_\mu n_{\nu'} n . \Gamma \Lambda %\non \\ &&- A_9 \Big( \Gamma_\mu 
% \Lambda \Gamma_{\nu'} -\sqrt{\frac{u}{u+2}}\Gamma_\mu n . \Gamma \Lambda 
%\Gamma_{\nu'} \Big), 
\eea

where we have used  the relation\footnote{We drop constants of integration as
the coefficients have to vanish as $u \rightarrow \infty.$}
\be F  = G = \int_\infty^u du' \frac{1}{\sqrt{u' +2}} \Big( A_9' 
- \frac{A_9}{2(u' +2)} \Big) = \frac{A_9}{\sqrt{u+2}}. \ee

Thus using four of the relations in \C{group1}, we see that the propagator 
depends on only four of the ten independent structures. 
Finally using the relations involving $F_1$ and $F_3$ in \C{group1}, we see that
\bea &&A_1 g_{\mu\nu'} \Lambda  = 
-{\cal{D}}_{\mu} \Big( \sqrt{u(u+2)} A_1 n_{\nu'} \Lambda  
+ u A_1  n_{\nu'} n .\Gamma \Lambda \Big) \non \\&&
+u \sqrt{u+2} H' n_\mu n_{\nu'} 
\Lambda + u^{3/2} H' n_\mu n_{\nu'} n. \Gamma \Lambda - \frac{\sqrt{u}}{u+2} H
g_{\mu\nu'} n .\Gamma \Lambda,
\eea
where 
\be H = \sqrt{u+2} A_1.\ee

Again neglecting the ${\cal{D}}_\mu (\ldots)$ and ${\cal{D}}_{\nu'} (\ldots)$ terms, this 
leads to
\bea \label{only5}
\Theta_{\mu\nu'} (z,w) &=&  \Big( (d+1) A_8 +u(u+2) A_1' +\frac{uA_1}{2} \Big)
n_\mu n_{\nu'} \Lambda + \Big( A_3 - \sqrt{\frac{u}{u+2}} A_1 \Big) 
g_{\mu\nu'} n . \Gamma \Lambda 
\non \\ &&+
\Big( A_4 + u^{3/2} \sqrt{u+2} A_1' + \frac{u^{3/2} A_1}{2\sqrt{u+2}} \Big) 
n_\mu n_{\nu'} n . \Gamma \Lambda  
\eea

Note that using the remaining relations involving $F_5$ and $F_7$ in \C{group1}, we cannot
remove any other structure, thus showing that three terms remain. In fact, it is 
easy to check that
the coefficient of the $n_\mu n_{\nu'} \Lambda$ term in \C{only5} identically vanishes, thus
leaving only two different structures. 
This gives the gauge--invariant
part of the propagator, and any other propagator is related to it by pure gauge terms.
Note that we could have chosen to keep any subset of the structures in \C{gravitinoprop}, however 
it seems to us that the choice in \C{only5} is convenient for calculating correlators 
involving a bulk gravitino exchange in $AdS_{d+1}$.

Thus using the relations \C{changebas2}, the gauge invariant part of the massless
gravitino propagator is given by  
\bea \label{gravpropdef}
\Theta_{\mu\nu'} (z,w) &=&  A (u)
\p_\mu \p_{\nu'} u (\p u .\Gamma) \Lambda + B (u) \p_\mu u \p_{\nu'} u (\p u .\Gamma) \Lambda \eea 
where 
\bea \label{valfour} A (u) &=& -2^{d +3/2} \lambda (d+1) (u+1)  
u^{-(d+3)/2} (u+2)^{-d/2 -2}
, \non \\ 
B (u) &=& 2^{d + 3/2} \lambda (d+1) u^{-(d+5)/2} (u+2)^{-d/2 -3} \Big(
(d+2) (u+1)^2 +1 \Big) \eea 
and $\lambda$ is given by \C{vallambda}.

\section{The spinor parallel propagator}

Now in explicit calculations of 
correlators involving the bulk gravitino exchange in Euclidean $AdS_{d+1}$ 
we also need the expression for the spinor parallel propagator 
$\Lambda^\alpha_{~\beta'} (z,w)$, which we 
now deduce. We work in Euclidean $AdS_{d+1}$ defined as the upper half space in
$z^\mu \in \mathbf{R}^{d+1}$, with $z^{0} > 0$, and metric given by
\be
ds^{2}=\sum_{\mu,\nu=0}^{d} g_{\mu\nu}dz^{\mu}dz^{\nu}
=\frac{1}{z_{0}^{2}}(dz_{0}^{2}+\sum^{d}_{i=1}dz_{i}^{2}). \ee
The coordinates $z_\mu$ will be raised and lowered with the
flat space metric unless otherwise 
mentioned. We choose the vielbein to be given by
\be e_{\mu}^{a}=\frac{1}{z_{0}}\delta_{\mu}^{a}, \ee
so that the spin connection is given by
\be w_{\mu}^{ab}=\frac{1}{z_0}
(\delta^{a}_{0}\delta^{b}_{\mu}-\delta^{b}_{0}\delta^{a}_{\mu}),\ee
where $a,b$ are tangent space indices. Thus the Dirac matrices in 
curved space $\Gamma^\mu$ will be related 
to those in the tangent space $\gamma^a$ by the relation
$\Gamma^\mu =e^\mu_a \gamma^a$, where  
$\{ \gamma^a, \gamma^b \} =2\delta^{ab}$. Note that in this coordinate system 
the chordal distance $u$ is given by
\be u =\frac{(z-w)^2}{2z_0 w_0}. \ee 

To deduce the expression for $\Lambda^\alpha_{~\beta'}$ that is useful for
explicit calculations, we shall use the known 
expression for the bulk--to--bulk spinor propagator. We shall equate 
the expressions for the
propagator given in~\cite{Kawano:1999au} and~\cite{Muck:1999mh} to obtain an expression
for $\Lambda^\alpha_{~\beta'}$. It should be noted that the two expressions 
involve non--trivial
factors of the mass $m$ of the spinor, and these factors 
cancel in the expression for $\Lambda^\alpha_{~\beta'}$, 
as they should. The overall numerical factor of $\Lambda^\alpha_{~\beta'}$ is fixed 
because $\Lambda^\alpha_{~\beta'} (z,z)=\delta^\alpha_{~\beta'}$. 
We now briefly outline the steps of our 
analysis, where we denote the spinor propagator by $S (z,w)$.       

From~\cite{Kawano:1999au}, we see that 
the expression for the spinor propagator is given by
\be \label{spinprop}
S (z,w) = -\frac{1}{\sqrt{z_0 w_0}} \Big[ (\gamma^\mu z_\mu P_- - P_+ 
\gamma^\mu w_\mu) G_{\Delta_-} (u)
+ (\gamma^\mu z_\mu P_+ - P_- \gamma^\mu w_\mu) G_{\Delta_+} (u) \Big],\ee 
where
\be G_{\Delta} (u) = -\frac{\Delta \Gamma (\frac{\Delta}{2}) 
\Gamma (\frac{\Delta +1}{2})}{4\pi^{(d+1)/2}
\Gamma (1+\Delta -\frac{d}{2}) (1+u)^{\Delta +1}} F \left(
\frac{\Delta}{2} +1, \frac{\Delta +1}{2} ; 1+\Delta 
-\frac{d}{2} ; \frac{1}{(1+u)^2} \right),\ee
\be \Delta_\pm = \frac{d}{2} + m \pm \frac{1}{2},\ee
and
\be P_\pm = \frac{1 \pm \gamma_0}{2}.\ee
Using the relations among hypergeometric functions~\cite{Abramowitz}
\bea F \left( \alpha, \beta; \gamma; x \right) &=& (1-x)^{\gamma-\alpha-\beta} 
F (\gamma -\alpha, \gamma -\beta; \gamma; x) , \non \\
F \left(\alpha, \alpha+\frac{1}{2}; \gamma; x \right ) &=& 
(1 \pm \sqrt{x})^{-2\alpha} F \left( 2\alpha, \gamma- \frac{1}{2}; 2\gamma -1; 
\frac{\pm 2\sqrt{x}}{1 \pm \sqrt{x}} \right),\eea
we see that \C{spinprop} becomes
\bea \label{moreeqn}
S (z,w) = \frac{1}{2^{m +(d+1)/2} \pi^{d/2}} \frac{\Gamma (m +\frac{d+1}{2})}{\Gamma (m 
+\frac{1}{2})} \frac{1}{{\sqrt{z_0 w_0}} (u+2)^{m+(d+1)/2}} \non \\ \times
\Big[ (\gamma^\mu z_\mu P_- - P_+ \gamma^\mu w_\mu) F \left( m +\frac{d+1}{2} , m ; 2m; 
\frac{2}{u+2} \right) \non \\ +\frac{m +\frac{d+1}{2}}{(2m+1)(u+2)} 
(\gamma^\mu z_\mu P_+ - P_- \gamma^\mu w_\mu) 
F \left( m +\frac{d+3}{2} , m+1 ; 2m +2; \frac{2}{u+2} \right) \Big]. \eea 
On using the relations
\bea \gamma F \left( \alpha, \beta; \gamma; x \right) -\gamma F \left( \alpha, 
\beta+1; \gamma; x \right) + \alpha x F \left( \alpha +1, 
\beta +1; \gamma +1; x \right) =0, \non \\
\gamma F \left( \alpha, \beta; \gamma; x \right) - (\gamma -\beta)  F \left( 
\alpha, \beta; \gamma +1; x \right) - \beta  F \left( \alpha, 
\beta +1; \gamma +1; x \right) =0,\eea
we finally get from \C{moreeqn}
\bea \label{spinprop1}
S (z,w) = -\frac{1}{2^{m +(d+3)/2} \pi^{d/2}} \frac{\Gamma (m +\frac{d+1}{2})}{\Gamma 
(m +\frac{1}{2})} \frac{1}{{\sqrt{z_0 w_0}} (u+2)^{m+(d+1)/2}}  \non \\\times
\Big[ (\gamma^\mu z_\mu \gamma_0 + \gamma_0 \gamma^\mu w_\mu) 
F\left( m +\frac{d+1}{2} , m ; 2m +1; \frac{2}{u+2} \right) \non \\
- \gamma^\mu (z -w)_\mu F \left( m +\frac{d+1}{2} , m+1 ; 2m +1; 
\frac{2}{u+2} \right)\Big]. \eea
We now consider the expression for the spinor propagator given in~\cite{Muck:1999mh}
constructed using the geometric objects we have mentioned before. There
are two expressions for the propagator depending on the sign of $m$, 
and for our purpose of
matching the answer we choose the 
positive sign which is allowed for generic choices of $m$, which 
we also take to be positive.
This gives
\bea \label{spinprop2}
S (z,w) = -\frac{1}{2^{2m +d+1} \pi^{d/2}} \frac{\Gamma (m +\frac{d+1}{2})}{\Gamma 
(m +\frac{1}{2})} \left( \frac{2}{u+2} \right)^{m+d/2}  \non \\ \times
\Big[ F \left( m +\frac{d+1}{2} , m ; 2m +1; \frac{2}{u+2} \right) \non \\
- \frac{\p_\mu u}{u+2} \Gamma^\mu F \left(m +\frac{d+1}{2} , m+1 ; 2m +1; 
\frac{2}{u+2}\right)\Big] \Lambda (z,w). \eea
Equating \C{spinprop1} and \C{spinprop2} we get
\be \label{defL}
\Lambda (z,w) = \frac{\gamma^\mu z_\mu \gamma_0 + \gamma_0 
\gamma^\mu w_\mu}{\sqrt{2w_0 z_0 (u+2)}} 
,\ee
which satisfies $\Lambda (z,z) =1$. 
Note that \C{defL} has a very simple form and one could have directly guessed the 
answer.\footnote{The other obvious
choice $\widetilde\Lambda (z,w) = \frac{1}{\sqrt{2w_0 z_0 (u+2)}} 
(\gamma^\mu w_\mu \gamma_0 + \gamma_0 \gamma^\mu z_\mu)$ does not satisfy 
$n^\mu D_\mu \widetilde\Lambda (z,w)=0$. In fact $\widetilde\Lambda (z,w) =\Lambda (w,z)$.}
As a consistency check, we note that
\bea 
D_\mu \Lambda (z,w) &=& -\frac{\p_\mu u}{2(u+2)} \Lambda (z,w) +\frac{\gamma_\mu
\gamma^\nu (z -w)_\nu}{2z_0 \sqrt{2z_0 w_0 (u+2)}} \non \\
&&= \frac{1}{2} \sqrt{\frac{u}{u+2}} (\Gamma_\mu
\Gamma^\nu n_\nu -n_\mu) \Lambda (z,w),\eea
which leads to $n^\mu D_\mu \Lambda =0$, which is one of the 
defining properties of $\Lambda$.

So the gauge invariant part of the massless gravitino propagator in $AdS_{d+1}$ is given by
\C{gravpropdef}, \C{valfour} and \C{defL}. It would be interesting to derive the
propagator using other approaches and check the 
gauge invariant part with the results we have obtained.   

\appendix
\section{The flat space propagator}

One needs to construct the flat space propagator in order to fix the overall
normalization of the coefficients in the propagator in $AdS_{d+1}$. 
In flat space, substituting \C{gravitinoprop} into \C{rel3} yields the coupled set
of equations
\bea \label{twoeqns} 
A_4' +\frac{d+2}{\mu} A_4 -mA_2 =0 , \non \\
A_2' -\frac{2}{\mu} A_2 -mA_4 =0,
\eea
where $'$ denotes derivative with respect to $\mu$, which has the solution
\be A_2 (\mu) = C (m\mu)^{(1-d)/2} K_{(d+3)/2} (m\mu), \quad 
A_4 (\mu)= -C (m\mu)^{(1-d)/2} K_{(d+5)/2} (m\mu) ,\ee
on demanding singular behaviour at $\mu =0$, where we have used the recursion relation
\be \frac{d K_\nu (z)}{dz} = \frac{\nu}{z} K_\nu (z) -K_{\nu +1} (z).\ee
Now the set of equations \C{twoeqns} does not receive 
any contact term contributions and so $C$
remains undetermined. In order to determine $C$, we want to 
consider those equations which have 
delta functions on the right hand side. Generally one would expect to obtain two coupled
differential equations in two variables which one can solve easily, 
but it turns out not to be the 
case.  

So to determine $C$ we consider one of the equations coming from inserting 
\C{gravitinoprop} into 
\C{rel3} given by 
\be \label{maineq}
A_3' +\frac{d}{\mu} A_3 -\frac{2}{\mu} A_6 -m A_1 = -\delta^{d+1} (z-w).\ee
In order to determine $C$ the standard technique is to integrate both sides 
and pick up the contribution of the left--hand side at the origin. The main motivation for 
considering \C{maineq} is that among all the equations this is the only equation 
where the derivative acts on only one function, there is a non--vanishing
delta function contribution and the 
first two terms on the left--hand side are of the form
\be f' +\frac{d}{\mu} f. \ee   
Now for $f= g'$, we know that $g'' + \frac{d}{\mu} g' = \Box g$, and 
this fact will be useful to 
us in determining $C$. So first we need to obtain the expressions for the coefficients in 
\C{maineq} which are solutions of the homogeneous equations for flat space. 
Using the relation
\be \label{nocont}
A_3 = \frac{1}{2} \Big( A_4 + (d+1) A_6 \Big), \ee  
which follows from inserting \C{gravitinoprop} into \C{rel1}\footnote{In fact 
there are no contact term contributions to \C{nocont}.}, we see 
that \C{maineq} reduces to
\be \label{fixcont}
\Big( A_4' + (d+1) A_6' \Big) +\frac{d}{\mu} \Big( A_4 +(d+1) A_6 \Big) 
-\frac{4}{\mu} A_6 -2m A_1 = -2\delta^{d+1} (z-w). \ee

In order to analyze \C{fixcont}, apart from $A_4$, we also need the expressions 
for the coefficients $A_1$ and $A_6$ which are the solutions of the homogeneous system of 
equations. In fact $A_1$ and $A_6$ are given by  

\bea \label{A1}
A_1 &=& \frac{d}{d-1} A_2 +\frac{(d+1)(d+3)}{m^2 \mu^2} A_2 + \frac{d+1}{m\mu} A_4
\non \\ &&=\frac{d}{d-1} C (m\mu)^{(1-d)/2} K_{(d+3)/2} (m\mu) - (d+1) C 
(m\mu)^{-(d+1)/2} K_{(d+1)/2} (m\mu),\eea
and
\bea \label{A6}
A_6 &=& \frac{1}{d-1} \Big( A_4 + \frac{2(d+1)}{m\mu} A_2 \Big) \non \\ 
&&= \frac{C}{d-1} (m\mu)^{(1-d)/2} \Big( \frac{2(d+1)}{m\mu} K_{(d+3)/2} (m\mu) 
- K_{(d+5)/2} (m\mu) \Big),\eea 
where we have used the recursion relation
\be K_\nu (z) = K_{\nu -2} (z) +\frac{2(\nu -1)}{z} K_{\nu -1} (z). \ee

Now consider integrating both sides of \C{fixcont} over a small sphere centered at the
origin $\mu =0$, which produces a factor of $\int d\mu \mu^{d}$ in the integrand. 
Using the expression at small $z$ for $\nu$ positive,
\be K_\nu (z) = \frac{1}{2} \Gamma (\nu) \Big( \frac{z}{2} \Big)^{-\nu}, \ee
it is easy to see that $A_1, A_4$ and $A_6$ all contribute at the origin individually
to the integral. We want to group the various contributions together in such a way
that the constant $C$ can be determined easily. To do that we use the combination 
\be {\widetilde{A}} = A_4 + (d+3) A_6, \ee 
and it is easy to see that 
\be \label{defA}
{\widetilde{A}} = -\frac{2(d+1)C}{(d-1)} (m\mu)^{(1-d)/2} 
K_{(d+1)/2} (m\mu).\ee
Then we rewrite \C{fixcont} as
\be \label{fixcont2}
\Big( {\widetilde{A}}'  +\frac{d}{\mu} {\widetilde{A}} \Big) 
-2 \Big( A_6' +\frac{d+2}{\mu} A_6 +m A_1 \Big) = -2\delta^{d+1} (z-w). \ee

Now consider the second set of terms in \C{fixcont2} which is proportional to
\be \label{cont2}
\frac{d A_6 (z)}{dz} +\frac{d+2}{z} A_6 (z)+ A_1 (z),\ee
where $z =m\mu$. Now when integrating \C{fixcont2} over the small sphere centered 
at the origin and looking at the contribution due to \C{cont2}, we see that
the individual 
terms $A_6'$ and $A_6/z$ both contribute to the 
integral at $O(z^{-d-3})$ and $O(z^{-d-1})$, 
while the $A_1$ term contributes at $O(z^{-d-1})$. In order to see this, we
use the expressions \C{A1} and \C{A6}, and the expansion of the Bessel function
at small $z$ given by
\be K_\nu (z) = \frac{1}{2} \Gamma (\nu) \Big( \frac{z}{2} \Big)^{-\nu} \Big[ 1 
+\frac{z^2}{4(1-\nu)} +\frac{z^4}{32(1-\nu)(2-\nu)}   \Big], \ee
for $\nu= (d+1)/2,(d+3)/2$ and $(d+5)/2$. However the total contribution due to
\be \frac{d A_6 (z)}{dz} +\frac{d+2}{z} A_6 (z) \ee
vanishes at $O(z^{-d-3})$, and gives
\be \label{partcont} 
-\frac{2^{(d+1)/2} C}{d-1} \Gamma \Big( \frac{d+3}{2} \Big)\ee
at $O(z^{-d-1})$. Now \C{partcont} is exactly cancelled by the contribution due to
$A_1$ at $O(z^{-d-1})$, and so \C{cont2} does not contribute to the integral at the 
origin. Thus writing ${\widetilde{A}} = d\Sigma/d\mu$ and integrating over the small sphere, 
\C{fixcont2} gives 
\be \Omega_{d+1} {\rm lim}_{\mu \rightarrow 0} \mu^{d} \frac{d\Sigma}{d\mu} = 
\Omega_{d+1} {\rm lim}_{\mu \rightarrow 0} \mu^{d} {\widetilde{A}}  = -2,\ee
where $\Omega_{d+1} = 2\pi^{(d+1)/2}/\Gamma((d+1)/2)$. So using \C{defA} we get that
\be \label{valC}
C = \frac{(d-1) m^d}{(2\pi)^{(d+1)/2}(d+1)} ,\ee
thus giving the flat space propagator. 
The expression for $A_9$ is given by
\be  \label{expA92}
A_9 =  C (m\mu)^{-(d+1)/2} K_{(d+1)/2} (m\mu) - \frac{C}{d-1} 
(m\mu)^{(1-d)/2} K_{(d+3)/2} (m\mu),\footnote{Note that $A_9 \sim \mu^{-(d+1)}$ as
$\mu \rightarrow 0$, which is consistent with the massive propagator.} \ee
which is also needed.

\section*{Acknowledgements}
We are grateful to M.~B.~Green for various comments and enlightening discussions. We would also 
like to thank 
D.~Z.~Freedman, R.~Maier, J.~Maldacena, K.~Okuyama, and L.~Rastelli for useful 
comments. The work of A.~B. is supported in part by NSF Grant No. 
PHY-0503584. L.~I.~U. would like to thank 
CONACyT Mexico and the ORSAS Scheme for financial support.

%\newpage
%\bibliographystyle{amsunsrt-es}
%\bibliography{myrefs}
%\bibliographystyle{utphys}
%\bibliography{myrefs}

%\end{thebibliography}
%\end{document}

%\begin{thebibliography}{10}

%\ifx\undefined\bysame
%\newcommand{\bysame}{\leavevmode\hbox to3em{\hrulefill}\,}
%\fi

\providecommand{\href}[2]{#2}\begingroup\raggedright\endgroup

\end{document}